\documentstyle[12pt]{article}

\title{\bf Nonachromaticity and reversals of topological phase as a
function of wavelength }

\vspace{40mm}

\author{\bf Rajendra~Bhandari}

\date{ }

\begin{document}

\maketitle
\vspace{1mm}
\begin{center}
\begin{tabular}{ll}
            & Raman Research Institute, \\
            & Bangalore 560 080, India. \\
            & email: bhandari@rri.ernet.in\\
\end{tabular}
\end{center}
\vspace{10mm}

\begin{abstract}

Contrary to the property of achromaticity (independence of wavelength)
usually associated with topological phases, we describe conditions 
under which topological phases encountered in optics can show 
sharp changes and reversals for small changes in wavelength, a 
phenomenon originating in the occurrence of phase singularities,
earlier observed in interference experiments.


\end{abstract}

\newpage
\noindent

Achromaticity of the topological phase arising from 
polarization transformations of light \cite{panch1}
has been the basis of several applications \cite{rbreview}.
Achromatic retarders based on Pancharatnam's work are routinely 
used in Astronomical polarimetry. Using the standard example 
of the ``QHQ retarder", where Q and H are quarterwave and 
halfwave retarders, 
the topological origin of achromaticity was explained in 
ref.\cite{qhqjumps} where it was also shown that at wavelengths 
far removed from the design wavelength $\lambda_0$ of the QHQ retarder 
it can show the opposite behaviour i.e. sharp changes and reversals 
of phase retardation originating in phase singularities \cite{jumps,rbdirac,iwbs}.
Direct interferometric observations of phase singularities were 
reported in \cite{rbdirac,iwbs}. 
The new result I wish to report here is that such sharp changes 
and reversals of phase retardation of a QHQ retarder can in fact be made to 
occur at wavelengths arbitrarily close 
to $\lambda_0$. This would happen if the retardation of 
Q and H were equal to $(2n+1/2)\pi$ and $(4n+1)\pi$ respectively,
where n is an integer which could be made large.

A simple way to understand the effect is to imagine a standard 
two-slit interference experiment in which fringes are formed on a screen placed some 
distance away from the slits.  A QHQ retarder is placed in the path 
of one of the two beams. Let the eigenstates of the retarder at 
$\lambda_0$ be $\mid X>$, i.e. linear polarization along the x-direction.
For an $\mid X>$-polarized source illuminating the slits,
the intensity on the screen, by Pancharatnam's prescription, is given by 
\begin{equation}
I = I_1 + I_2 + 2 \sqrt(I_1 I_2) Re(<X \mid U \mid X>e^{i\phi}),
\end{equation}
where U is the 2x2 unitary polarization transformation matrix 
representing the QHQ device and $\phi$ is a path-dependent phase 
varying along the screen. U is a function of two parameters, 
(i) the retardation $2\delta$   of Q (hence $4\delta$ of H) 
at the wavelength $\lambda$, assuming a dependence 
$\delta=(\lambda_0/\lambda)(\pi/4)$, 
(ii) $\rho$, the orientation of the fast axis of H with respect to 
the X-direction. 
The modulus and phase of the complex visibility function 
$\gamma = <X \mid U \mid X>$ determine respectively the fringe 
contrast and the phase shift caused by the QHQ device. 
The assumed $\lambda$-dependence of the retardation results 
when the two refractive indices of Q and H are independent 
of wavelength.
It was shown in ref.\cite{qhqjumps} that (1) when $2\delta=(2m+1)(\pi/4)$,
which happens when $\lambda=(2\lambda_0)/(2m+1)$, 
phase singularities occur at the HWP orientations $\rho=m\pi$;
m being an integer. 
At these values, the modulus of $\gamma$ goes to zero leading to 
zero contrast of the fringes and its  
phase undergoes a sharp jump of magnitude $\pi$. In a closed circuit 
around any one of these singularities, the phase changes by $\pm 2\pi$.
Such phase changes have been verified experimentally in interference 
experiments \cite{rbdirac,iwbs}

(2) When $2\delta=2m\pi-3\pi/2$, which happens when 
$\lambda=\lambda_0/(4m-3)$, the visibility is constant in magnitude 
and its phase increases linearly with $\rho$, the device acting as 
a pure phase retarder.
(3) When $2\delta=2m\pi-\pi/2$, which happens when 
$\lambda=\lambda_0/(4m-1)$, the visibility is constant in magnitude 
and its phase decreases linearly with $\rho$, the device acting as 
a pure phase retarder with the opposite sign.

The key point of this paper is that if the retardation of Q and H were 
chosen to be $2n\pi+\pi/2$ and $4n\pi+\pi$ respectively (multi-order
waveplates), the spacing in $\lambda$ at which singularities and hence 
the phase reversals occur can be made arbitrarily small. 
If we assume again a $1/\lambda$ dependence of retardation of Q and H,
the condition for phase reversal closest to the design wavelength 
for such a retarder is given by, 
$(2n\pi+\pi/2)(\lambda_0/\lambda)=(2n\pi+3\pi/2)$. 
This gives, for $n=1$, $\lambda=.71\lambda_0$;
for $n=2$, $\lambda=.82\lambda_0$; for $n=3$, $\lambda=.87\lambda_0$
and so on.  It can be concluded therefore that while 
topological phases {\it can} be made achromatic which can be 
useful, they are not necessary so.

It may also be pointed out that there are other contexts in 
interferometry wherein $\pi$ phase jumps 
accompanying zero crossings of a real visibility function can be seen as 
manifestations of a phase singularity of a complex visibility function 
in a higher dimensional space. In such an enlarged description, 
the $\pi$ phase shift acquires a sign that is measurable. 
We wish to cite one such example.

In an interesting extension of Pancharatnam's phase criterion to the 
physics of interference of mixed states, Sj\"{o}qvist et.al.
\cite{rhoui} have shown that if a beam of particles with internal 
degrees of freedom , 
in a mixed state with a density matrix ${\rho}_0$,
is split in a Mach-Zhender interferometer and a unitary transformation 
$U_i$ acts on the space of N internal states in one of the two paths, 
a phase difference given by arg Tr($U_i$${\rho}_0$) is introduced 
between the two beams. Here ``phase difference" is defined as 
the shift in the maximum of the interference pattern. This quantity, 
which is measurable, reduces to the phase shift 
arg($<{\psi}_0 {\mid U_i \mid} {\psi}_0>$), defined by 
Pancharatnam \cite{panch1} for pure states $\mid{\psi}_0>$.
We wish to point out that the mixed state phase
as defined in ref.\cite{rhoui} becomes indeterminate at points in the 
parameter space where ${\mid Tr(U_i{\rho}_0)\mid}$=0. Eqn.(8) 
in the paper shows that at such points the interference pattern is 
uniform, with no fringes. Just as in the pure state case discussed above, 
such points constitute singularities of 
phase involving discontinuous phase jumps and, depending upon 
the context, may occur as  singular lines or surfaces. 
The mixed state phase involves singularities in new kinds of parameter 
spaces which include variables representing decoherence of quantum 
states or depolarization in case of polarization states of light. 
One can therefore have measurable $2n\pi$ topological phases resulting 
from cycles in parameter spaces consisting of degree of polarization 
and parameters of unitary transformation of polarization states.
Let us note that in the entire discussion above, we have been concerned 
with the total phase and not just a geometric part of the phase.

\end{document}